# PRECISE POSITIONING SYSTEMS FOR VEHICULAR AD-HOC NETWORKS[1]

Samir A. Elsagheer Mohamed[1&3], A. Nasr[2&3], Gufran Ahmad Ansari[3]

[1]Electrical Engineering Department, Faculty of Engineering, South Valley University, Aswan, Egypt.
Emails: samhmd@qu.edu.sa; samirahmed@yahoo.com

[2]Radiation Engineering Dept., NCRRT, Atomic Energy Authority, Egypt.
Email: Ashraf.nasr@gmail.com

[3]Currentlly affiliated to the College of Computer, Qassim University, P.O.B 6688, Buryadah 51453, Qassim, KSA.
Email: ansariga1972@gmail.com

## ABSTRACT

*Vehicular Ad Hoc Networks (VANET) is a very promising research venue that can offers many useful and critical applications including the safety applications. Most of these applications require that each vehicle knows precisely its current position in real time. GPS is the most common positioning technique for VANET. However, it is not accurate. Moreover, the GPS signals cannot be received in the tunnels, undergrounds, or near tall buildings. Thus, no positioning service can be obtained in these locations. Even if the Deferential GPS (DGPS) can provide high accuracy, but still no GPS converge in these locations. In this paper, we provide positioning techniques for VANET that can provide accurate positioning service in the areas where GPS signals are hindered by the obstacles. Experimental results show significant improvement in the accuracy. This allows when combined with DGPS the continuity of a precise positioning service that can be used by most of the VANET applications.*

## KEYWORDS

*Vehicular Ad-Hoc Networks; Positioning Systems; Neural Networks; Wireless Networks; Received Signal Stength.*

## 1 INTRODUCTION

The number of vehicles increases dramatically day after day. This in turn gives rise to the urgent need for regulation of vehicular traffic and improvement of vehicle safety on highways and urban streets. Recently, there have been intensive efforts to networked-intelligent vehicles to yield the safety enhancement, and other potential economic benefits that can result from enabling both communication between vehicles and vehicular feedback to an ad hoc network. This gives birth to what so called Vehicular Ad hoc Network (VANET), by which, intelligent vehicles can communicate among themselves and with road-side infrastructure [1]. VANET can provide many useful services, such as collision avoidance, cooperative driving, automatic driving, navigation and probe vehicle data that increase vehicular safety and reduce traffic congestion, and offer access to the Internet and entertainment applications [1-4].

However, there are several challenges in VANET yet to be tackled before the full deployment of VANET technology in all over the world. Among these challenges is the knowledge of the accurate vehicle location that can lead to many safety enhancing applications. It is clear that a

---

[1] This work is supported by the Scientific Research Deanship, Qassim University, KSA.

DOI : 10.5121/ijwmn.2012.4217                                              251

precise positioning system that allows each vehicle to know its location in real time and continuously is a must for VANET to be operational.

There exist several positioning techniques that are suitable for many applications [1][5-8]. The most famous one is the GPS (by using a set of satellites that feeds information about the position of a GPS receiver). However, all the existing positioning techniques including the GPS have several drawbacks. Lack of accuracy of the resulting measurements is the most unacceptable disadvantage. For example, GPS devices can produce an error of up to 50 meters [7]. This accuracy may seem to be acceptable for several applications. On the other side, other applications like collision avoidance, automatic driving and lane tracking demand precise and accurate positioning information. Therefore, the existing positioning techniques including GPS are not suitable for this kind of applications in VANET. Recently an enhancement to the GPS referred as the Differential GPS (DGPS) consisting in installing expensive ground stations can improve the accuracy significantly. However, GPS and DGPS and the similar techniques do not work in tunnels, undergrounds, and in high dense building areas, because the signal cannot be received or received very weak.

Some VANET services must have continuous and accurate positioning information that cannot be interrupted, otherwise great damage or malfunction can take place: example of such application is the collision avoidance.

In this paper, we propose positioning techniques that can be used in conjunction with DGPS to provide a continuous and robust real time localization service with the following features: precise; not expensive (no extra hardware is required except those already needed by the VANET infrastructure); reliable; work in all areas and work in real time.

The rest of this paper is organized as follows: In Section 2, the related works and research efforts are given. Descriptions of the proposed techniques are given in Section 3. Experimentations and the obtained results are given in Section 4. Finally, the Conclusions and the future works are given in Section 5.

## 2 RELATED WORKS

There exits two categories of positioning systems: indoors techniques and outdoor techniques. Indoors techniques [5] are not suitable for VANET due to its cost and the limited distance they support. We focus here on the outdoor techniques. The most famous one which is always mentioned in the research papers is the GPS [7][9]. However, GPS has several drawbacks as discussed below:

- Its accuracy, the civilian GPS has a limited accuracy in the order of 50 meters in each direction. This is really unsuitable of the most of the applications of VANET, .e.g. collision avoidance, automatic driving (lane tracking), etc. A survey of the most famous GPS device vendors, the best announced accuracy is +/- 5 meters, which gives an error of about 10 meters in each direction. They say that this accuracy is obtained in 95% of all the readings. The other 5% of the readings, the accuracy may be very bad. This still too much far away of being acceptable in the critical applications requiring a precise positioning system.

- In some places (e.g. inside tunnels, undergrounds, etc.), the GPS has no coverage. This is because the GPS receivers calculate the position based on the signals received from 4 simultaneous satellites. In such places, the received signals are too weak or no signals can be received at all. Therefore, the GPS receivers cannot calculate the position.

- GPS signals sent from the satellites can be simulated. In other words, there exist GPS simulators (devices generating fake GPS signals). As mentioned before, the real GPS signals are relatively weak signals. GPS simulators produce relatively strong signals and therefore, the GPS receiver will use the fake signals and consider the real signals background noise. As a result, the receiver will calculate the position based on the fake signals. Thus, any jammer equipped by this kind of GPS simulator can inject erroneous data that leads to make the vehicle believe that it is in completely different position.

The existing outdoor positioning techniques are as follows.
- Signal-strength-based techniques. The receiver calculates an estimate of its location based on the received signal strength from several wireless access points. Lack of accuracy and the access points must be placed with care to cover the roads. Reported results based on this technique shows poor accuracy [1][11][20]. In this paper, we will show how to improve the accuracy based on this technique.

- Time-Of-Arrival (TOA)-Based techniques. It is based on the travelled distance from the base station to the receiver of a known signal. This is the solution adopted by GPS [9]. As we mentioned before, its practical measurements are not acceptable for VANET, as it lacks the accuracy. In addition, the Signal cannot be received in some places (tunnels, undergrounds, and near dense areas of building). TOA techniques require a perfect synchronization between the clocks of the base stations and the receivers. This cannot be guaranteed except by using atomic clocks which are very expensive. Otherwise the accuracy is compromised.

- Techniques based on the round trip time [9]. The receiver sends a small packet to the base station and wait for a reply from it. The elapsed time is proportional to the distance between it and the base station. By some calculations the receiver can know its location provided that it can talk to at least three base stations each one knows its accurate location. Again the distance between the base stations and the receiver must be large to obtain good results.

- There exist several other works that focus on VANET [8-15]. Most of these works uses the Received Signal Strength and provide a framework for using the GPS. However, the lack of accuracy in the obtained results makes the methods unsuitable for some of VANET applications.

## 3 DESCRIPTION OF THE PROPOSED POSITIONING TECHNIQUES

In this Section we will outline the proposed positioning techniques.

### 3.1 Introduction

As previously mentioned GPS, as a positioning system, is not suitable for most of the VANET services. The main raison for that is its accuracy. To solve this problem, DGPS system is used in the case if the accuracy is of great importance. DGPS can give a precision of several centimeters depending on the quality of the used components. The idea is to use some ground base stations that knows their accurate positions and broadcast the correcting information to the other GPS receivers in their covering areas. The receivers obtain from the satellites the normal GPS signals (which are not accurate) plus the correcting information from the ground base stations. As a result, the receivers can know in real time their accurate position. It should be noted that DGPS cannot operate if the normal GPS signals cannot be received from the satellites. The ground base stations provides only correcting information in the area. As a known fact, GPS signals cannot be received in the undergrounds roads, in the tunnels, under the bridges, near tall buildings or objects, or in areas having many buildings (like the case of city centers).
Road-Side Unites (RSUs) are essential components in VANET. The functions of the RSUs are to: a) disseminate the information to nearby vehicles; b) collect information to the central offices; and c) to provide Internet connection and entertainment to passengers. These RSUs are wireless Base Stations, a.k.a. Wireless Access Point (WAP) or AP for short. They are used to broadcast and receive information from the integrated VANET controller in each vehicle: in VANET terminologies, the integrated controller is named as On-board Unit (OBU). The OBU has an ad-hoc wireless communication capability. It incorporates also the unit that provides the vehicle by its current locations in real time; we refer to this unit as the OBU Localization Unit (OBLU).

Therefore, the natural question is: why not we use these RSUs to provide private, cheap, and accurate positioning system for the vehicles? This also can work in all the areas that have VANET infrastructure installed including tunnels, undergrounds, etc.

In order to have a positioning system suitable to the requirements of VANET services and application, we propose the following. The OBLU must contain a GPS receiver capable of the DGPS error correction operation. This part of the OBLU will provide the positioning information accurately in all the areas where the GPS signals from satellites are received. In addition, this part detects if there are correcting signals received from the DGPS ground bases stations or not.

The second part of the OBLU will be able to provide the positioning information in the following cases: in areas where the GPS signals cannot be received from the satellites; or if these signals are very weak; or in the areas where the correcting information from the DGPS ground base stations cannot be received or received too weak. This part of the OBLU will use VANET infrastructure and the next methodology to provide the positioning information.

We propose to use the Received Signal strength (RSS) from the RSU (which is an essential part of VANET infrastructure). The use the RSS as a positioning system for the indoor application is not new [1][5][11][14]. However, the reported results in the published works make this unsuitable to VANET (sometimes more than 50 meters as accuracy [11]). One contribution of this Paper is the new ways by which the RSS can be used to significantly improve the accuracy of the positioning (See Sections 44.2.2, 4.3.4 and 4.4 for more details about these improvements techniques).

### 3.2 Generic architecture of VANET possitioning system

Based upon the above introduction, we propose the following generic architecture of VANET positioning system that has to be followed in order to get a universal positioning system.

1) If the OBLU detects good GPS signals and at the same time it receives correcting information from the DGPS ground base station, then current location information will be obtained from the DGPS. We refer to it as $P_{DGPS}$

2) Otherwise, the following technique will be used to obtain the current location information. We refer to it as $P_{RSS}$

    a) On the road side, there must be several RSUs which are simply a wireless access points (WAP), that periodically broadcast their existence by sending beacons (small messages) that can be received by the OBLU in the vehicle in the coverage area.

    b) These RSU are planted on the side of the street. Each RSU is installed on a specific position. The RSU broadcasts its position within the beacons massage (the exact global position: the longitude, the latitude and the altitude). Each RSU must operate on a channel that does not interfere with the nearby RSU.

    c) The OBLU can measure the received signal strength intensity from all the RSUs in real time. The OBLU must hear from at least two RSUs, if the 2D location of the vehicle is desired. In the case of 3D positioning, at least three access points must be heard.

    d) If OBLU hears from many RSUs, this can improve the accuracy. For example, in order to obtain the 2D position of the vehicle on the street, only two RSU are need. But if it receives from three, we find three readings. By taking the average of the three values, the accuracy will be improved. Selecting which RSU to be

used in the calculation is based on several conditions (See Section 4.5 for more detail).

e) The OBLU can determine its global positioning based on the RSS and the absolute position of each RSS. Details on this Step are given in Sections 4.3 and 4.4.

## 4 FIELD EXPERIMENTS TO MEASURE RSSS FROM WAP.

### 4.1 Experiments Description

Two field experiments are conducted in realistic conditions. VANET hardware is installed on a portion of a street. The RSUs are wireless access points (WAPs) operating on Wi-Fi 802.11g. The on-board unit is simply a Laptop having the software that can measure the RSS from the WAPs at any place in the covering area.

#### 4.1.1 The first experiment

Let's first describe the environment and the precautions that we take for this experiment. We have measured the signal each 5m from the starting point 0 until the ending point at 200m. The elevation of the access points is 110 cm, above the ground. Distance between the line of access points and the parallel line of the test line equals 7m. All the access points and mobile host operates on Channel 6. A diagram of the setup of this experiment is shown in Figure 1.

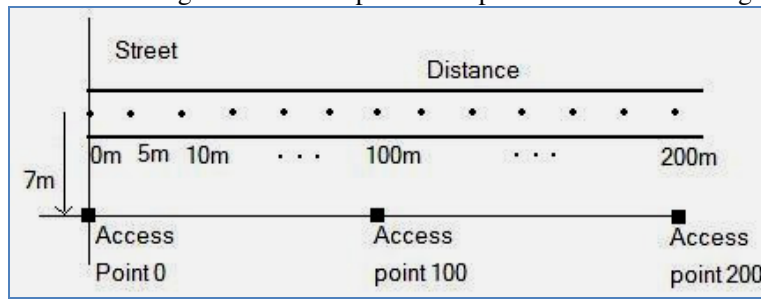

Figure 1. Experiment #1 Setup

#### 4.1.2 The Second Experiment

In this experiment our goal is to solve a fatal problem discovered in Experiment #1. Mainly, to reduce the Wi-Fi channel interference and hence increase the accuracy of the RSSI, each WAP operates on a non-overlapping channel (Chanel #1, 7, and 13). All other experiment setup is the same as the previous one.

### 4.2 Obtained Results

In this Section, we will present the obtained results for the two experiments. In all the Figures in this Section, the x-axis shows the distance on the street (the longitude displacement of the vehicle). The y-axis shows the value of the received signal strength intensity (RSS) from the wireless access point(s) WAPs. Remember, the RSS is negative and the far we are from the WAP, the less the RSS.

#### 4.2.1 Experiment #1 obtained results

Figure 2 shows the obtained RSS values from the three Access Points for the First Experiment (Experiment #1). From the shown data, we can see that the results are very bad, the data is very chaotic. It is expected that the RSS from any WAP decrease when we move far away of it (increasing the distance). However, for all the WAPs, this is not the case.

By inspecting the results, we can expect that no way to obtain accurate estimate of the position based on the RSSs. On other words, the accuracy will be very bad. That expectation is confirmed using the neural network as a learning tool. The accuracy is very bad, about 80 meters in any direction (for the lack of space, results are omitted).

A very interesting finding here is that this obtained result by the neural network (NN) for this experiment is almost the same as those found in the literature [9][11]. The published works for using the RSS as positioning technique suffer from the channel interference problem. Once we collected the results, we have thought in the problem. We, then, realized that all the WAPs operated on the same frequency channel. It is known that the Wi-Fi networks have 11-13 channels depending on the country. If all the WAPs use the same channel, that will cause a significant noise and interference. In such case, the estimation of the distance from the RSS is very difficult and hence the system gives inaccurate results. Thus, the interference problem has to be solved in order to improve the accuracy.

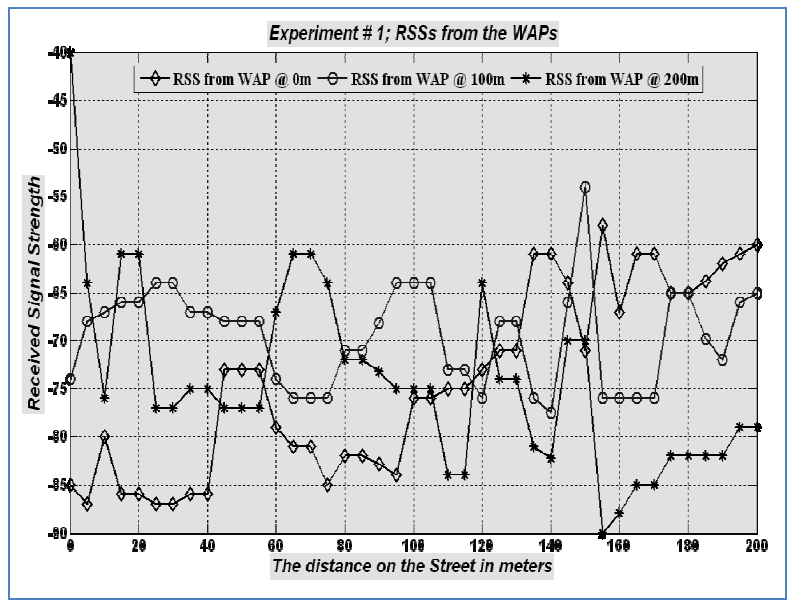

Figure 2. Obtained RSS from the three Access Points for the First Experiment (Experiment #1)

#### 4.2.1.1 How to solve the interference problem?

It is also known that the channels do not overlap if they are more than 5 channels apart. For example, if one WAP use channel #1, the second uses channel #6 and the third uses channel #11, then there is not overlap and thus no noise nor interference that can degrade the results. That is confirmed using the obtained results.

#### 4.2.2 Experiment #2 obtained results

Figure 3 depicts the obtained RSS values from the three Access Points for Experiment #2.

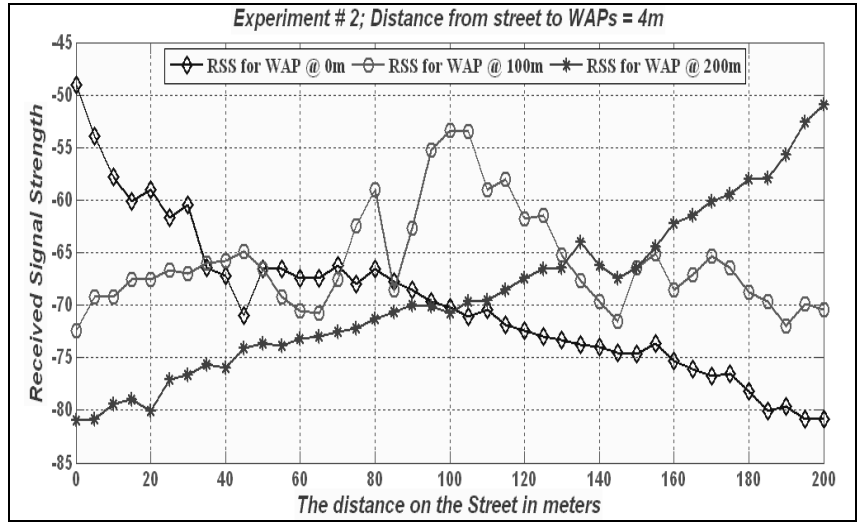

Figure 3. Obtained RSS from the three Access Points for the Second Experiment (Experiment #2).

### 4.2.3 Data Analysis of Experiment #2

As can be seen from Figure 3, once the interference of the channels is eliminated, the results are improved significantly. Moreover, one of the most important discoveries in this work is that the RSS from any WAP is still chaotic even if there is no interference if the distance to the WAP is less than 60 meters. By increasing the distance, we obtain better results. This can be seen for AP@0m and AP@200m. When the distance is greater than 60 meters, the RSS is inversely proportional to the distance (the expected behavior). For WAP@100m, this is also true. In the middle part of Figure 3, the data are chaotic. However, it starts to be correct at the boundaries (when the vehicle is far away from the WAP). This is a known problem in the wireless networks. It is referred as the near-far problem, where the interference between the emitter (the WAP) and the receiver is more if they are near to each other. The interference decays as the distance increase. By making use of this observation, the accuracy can be improved significantly. See Section 4.4.

In the next subsections we present the use of the neural networks and the curve fitting to estimate the position of the vehicle based on the received signal strengths and the use of the obtained experimental data.

### 4.3 A positioning system using the neural networks for VANET

In this Section, we present the use of the neural networks (NNs) as a learning tool to estimate the location of the vehicle on the road based on the RSS from the WAPs. The full theory and the description of the NN are beyond the scope of this paper.

### 4.3.1 Few words about the neural networks

The idea of the NN is inspired from the biological nervosystem. Similarly to the biological neuron, the artificial neural network learns by experience. In other words, a set of learning examples must be collected for the problem in hand. These samples are simply a set of the system inputs and their corresponding outputs.

Using training algorithms, the NN will learn the system behavior based on these examples. Once trained, the neural network will act as the system. For any new inputs, the NN will estimate the outputs based on the learning experience.

### 4.3.2 Description of the approach with the NN

We have used Matlab 2009a to do most of the calculations, and the Figures in this paper. In addition, there is a very nice, easy to use, and powerful toolbox in Matlab called the 'Neural Network toolbox'. The inputs to the NN are the RSS from the WAPs. The output is the location of the vehicle on the street or the required position. The NN needs to be trained, validated then tested. Thus, using Matlab, the whole learning example (the set of samples in the form RSS-Position pairs), is divided into three sets: one for training; the second for validation; and the third for testing. This is done automatically by Matlab. The ratios are 70% for training; 15% for validation; and 15% for testing. The training stops when the performance on the validation set increase. This is to avoid a known problem with the NN which is the overtraining.

For the NN, we use the Feedforward architecture that contains the input layer connected to the inputs; the output layer, which gives the estimate of the location based on the learning experience. In addition, there is the hidden layer. This layer can contain a given number of hidden neurons. Moreover, the random seed for the random generator which is used to assign the initial weights of the NN is very important. For the same samples, and the same NN, but different random seeds, the trained networks are completely different. Thus, we take this as a parameter for all the data.

A Matlab script to automate most of the work is developed. Then the outputs can be saved and restored at any time to know the best NN that gives the best performance.

### 4.3.3 Working with data obtained in experiment #1

We have run the script on the data obtained from experiment #1. A total of 56 different NN architectures are created; trained; tested and validated. The performance measures are also calculated on both the testing and the whole samples.

Figure 4 shows the difference between the actual data (distance) and those estimated by the trained NN for the best one. As we can see from the Figure, error on the positioning may reach up to 60m. This is very high and not suitable at all for VANET. This bad result is that found on the literature, see for example [9]. Actually, people used the technique without taking into account the effect of the interference problem resulting from using the same channel for all the WAPs. This is why the published results for the mechanisms using the RSSs are not encouraging.

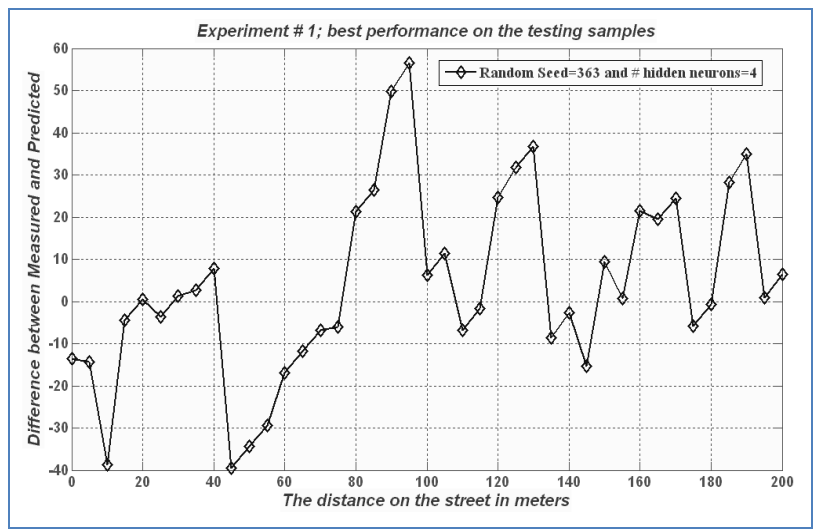

Figure 4. The performance of the best NN for experiment #1. The Figure shows the difference between the measured data experimentally and those calculated by the NN. The error can reach up to 60 meters. Correlation coefficients= 93 %

### 4.3.4 Working with experiment #2 data

Similar to what done in experiment #1, the whole work is repeated on the data for experiment #2. We have run the script on the data obtained from experiment #2. By using NNs with different hidden neurons and random seed generator: A total of 180 different NNs are trained, validated and tested. The performance measures are also calculated on both the testing and the whole samples. In Table 1, we show the obtained data sorted by the MSE on the whole samples and the max error obtained. In the table, we show also the standard deviation, the max error, the variance and the correlation for both the testing samples and the whole samples.

From this table, we can see that the best NN is obtained when the # of hidden neurons is 10 and for initial random seed =102. As expected the accuracy of this best one is very good compared to that of the first experiment. The MSE of the best NN is 6 against 172 for the first experiment. The standard deviation is about 3.2m, the max error is about 7.6 meters against 60 meters. Comparing these results to the GPS, we can see that we can reach accuracy better than the GPS. Thus, our technique can be used to complement the GPS in the areas where the GPS signal cannot be received or if it is very week.

Thus, the accuracy is improved significantly, by about a factor of 7. That is a huge improvement. This can also be seen from Figure 5 which depicts the difference between the actual data (distance) and those estimated by the trained NN.

### 4.3.5 Effect of removing the WAP@100m

We have removed the RSS from WAP@100m that we have thought may degrade the performance. The results show no further improvement (omitted for the sake of space). This is because three WAPs are better than two.

### 4.3.6 Using only one WAP but far away

We are seeking to improve the performance as much as possible. As described before, the near-far problem is a source of accuracy degradation. To examine its effect and the ability of the NN to learn, data from WAP @0m and WAP@200m only when the distance from each one is greater than 60m are used.

The obtained results also are not encouraging. Again, the NN cannot correctly learn from the few samples available for training. The next subsection presents another way to improve the performance, which is the curve fitting.

### 4.4 Positioning System Using Curve Fitting

In the previous Section, we have seen how to use the neural network for the positioning system. However, it is still in the range of 7 meters. We are seeking to improve this performance. We have explored the use of the curve fitting to obtain a polynomial equation that can be used to obtain the distance from the WAP given its RSS. Several types of curve fittings are evaluated using Matlab. The best result is given bellow. We aim to approximate the positioning problem according to the following equation. For the Linear model Polynomial of degree 4, we have the following equation:

$$\text{Distance} = p_1 R^4 + p_2 R^3 + p_3 R^2 + p_4 R + p_5,$$

where $R$ is the received signal strength.

Table 1. Results of exp. #2. By using NNs with different hidden neurons and random seed generator: A total of 180 different NNs are trained, validated and tested; only the top 5 are shown

| # | Hidden neurons | random Seed | Mean Square Error (MSE) | | Max absolute Error | | Standard Deviation | | Variance | | Correlation Coefficient | |
|---|---|---|---|---|---|---|---|---|---|---|---|---|
| | | | Testing Items | All Items | Testing Items | All Items | Testing Items | All Items | Testing Items | All Items | Testing Items | All Items |
| 1 | 10 | 102 | 8.4 | 6.0 | 5.4 | 7.6 | 3.2 | 2.5 | 10.0 | 6.1 | 0.997 | 0.999 |
| 2 | 8 | 102 | 15.9 | 7.5 | 8.0 | 8.0 | 4.2 | 2.7 | 17.4 | 7.2 | 0.997 | 0.999 |
| 3 | 8 | 1 | 5.9 | 7.9 | 3.5 | 8.8 | 2.4 | 2.8 | 5.9 | 8.1 | 1.000 | 0.999 |
| 4 | 4 | 899 | 8.2 | 8.1 | 5.4 | 8.6 | 2.9 | 2.9 | 8.5 | 8.3 | 0.999 | 0.999 |
| 5 | 9 | 363 | 38.6 | 8.3 | 12.4 | 12.4 | 5.5 | 2.8 | 30.1 | 8.1 | 0.998 | 0.999 |

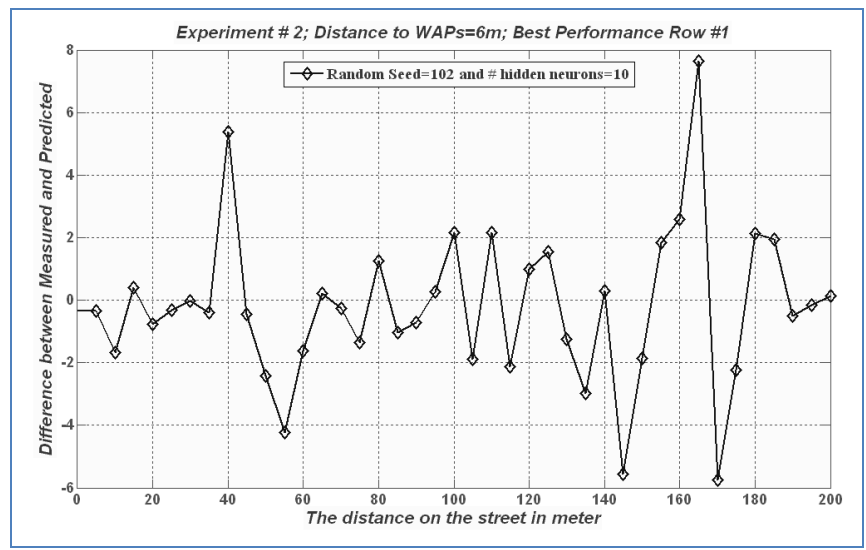

Figure 5. The performance of the best NN for experiment #2. The Figure shows the difference between the measured data experimentally and those calculated by the NN.

### 4.4.1 Starting from 60m

We have used the data from WAP@200m and by removing the data when the distance is less than 60m (the chaotic data). Again, this is to solve the near-far problem. Using Matlab "cftool" curve fitting tool, the estimated values of the Coefficients (with 95% confidence bounds) are as follows:

$p_1$= -0.005206, $p_2$= -1.553, $p_3$= -173.5, $p_4$= -8608, $p_5$= -1.601e+005.

The obtained Goodness of fit for the above data is: SSE: 422.7; R-square: 0.9917; Adjusted R-square: 0.9903; and RMSE: 4.197. The curve fitting results are shown in Figure 6.

We have calculated the distance for the given RSSs in the range. In Figure 7, we draw the difference between the actual data (measured) and the calculated ones. One can see that using only one WAP, the accuracy is almost the same as using the NN with three WAP. That is a very important finding. Staring from 60m solved the problem of the near-far problem that the wireless networks are suffering from.

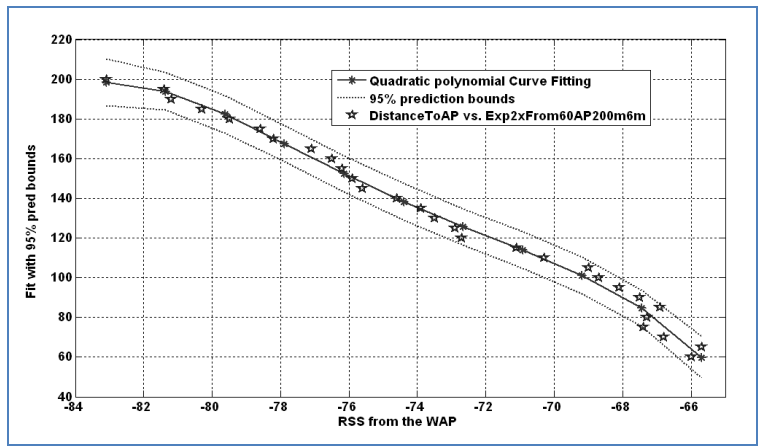

Figure 6. Curve Fitting results using cftool, when the Distance is greater than 60m

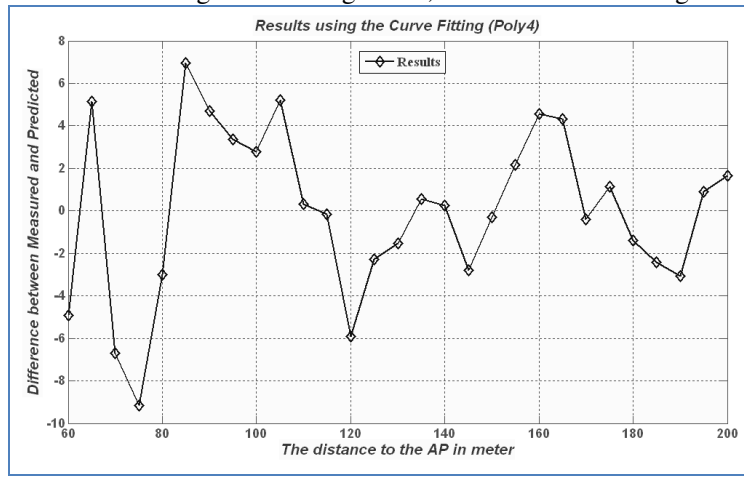

Figure 7 The performance of curve fitting for Exp # 2 by using only data from WAP@200m and excluding data before 60m. The Figure shows the difference between the measured data experimentally and those calculated by the NN.

### 4.4.2 Starting from 100m

From the previous section, we can see that the performance is bad at the beginning of the curve. However, it is getting better when the distance from the WAP increase. Thus, we redid the same experiment, but removing all the data when the distance is less than 100m. The results of the curve fitting are shown in Figure 8.

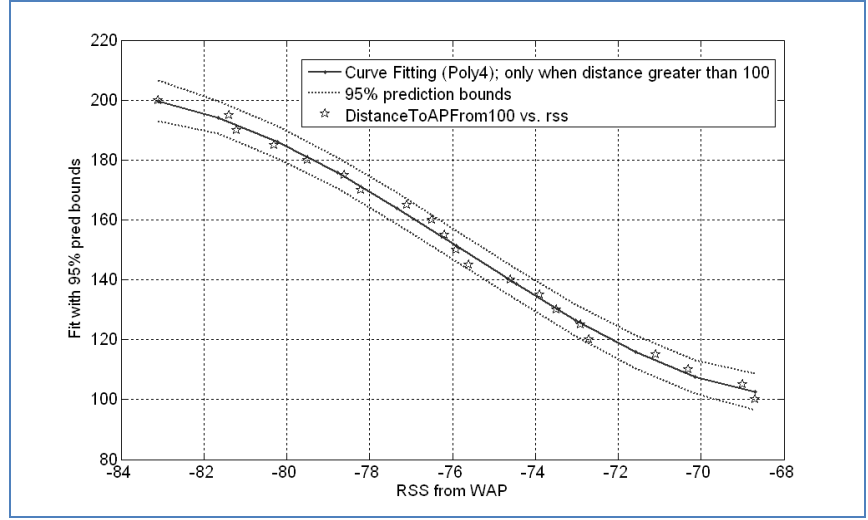

Figure 8. Curve Fitting results using cftool, when the Distance is greater than 100m

The coefficients of the polynomial equation are as follows: $p_1$=0.0004373, $p_2$=0.1736, $p_3$=24.4, $p_4$=1458, $p_5$ =3.171e+004. The obtained goodness of fit is: SSE=85.13, R-square=0.9956, Adjusted R-square=0.9945, RMSE= 2.307. Comparing these data with the ones in the previous experiment, we note a very significant improvement. This is also confirmed from Figure 9. We can see that the accuracy is improved (the max error is now 4 meters instead of 8 meters). This is by using only one WAP.

Again, if we inspect the figure carefully, we can see that the results when the distance is greater than 170m are within only 2 meters (A very accurate results). This result is much better that that obtained using the GPS. Moreover, it costs nothing (no extra hardware is needed, only the RSS from the RSUs), it can work in tunnels, undergrounds, and in the dense areas or near tall buildings where GPS cannot be used.

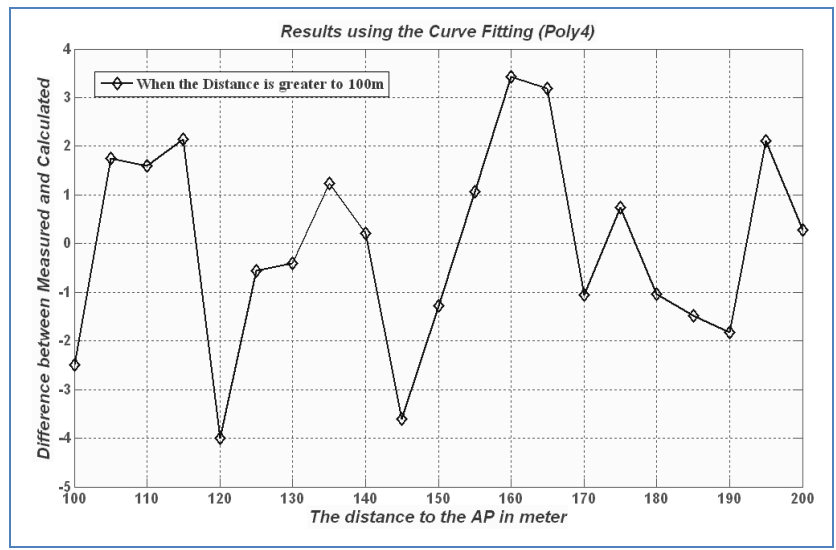

Figure 9. The performance of curve fitting for Exp #2 using only data from WAP@200m and excluding data before 100m. The Figure shows the difference between the measured data experimentally and those calculated by the NN.

## 4.5 The best approach for the positioning system

Based on all the results explained in the previous Subsections, we summarize here the best approach that can be used to build a system for an accurate positioning system using the RSSs from the RSUs for VANET. (See Section 3.2 for details on the generic system with conjunction of the DGPS.)

- The RSUs must be installed with 100m of minimum distance between each two.

- At any point on the road, the on-board unit must receive from at least two RSUs satisfying the next condition: these RSUs must be as far as possible (the less the RSS), and each operates on a different radio channel. This is to minimize the effect of the channel interference and the near-far problem.

- Using a simple triangulation technique, the 2D and the 3D coordinate point of the vehicle can be determined.

- If the on-board unit receives from more than two RSUs, this is very useful. That can be used to improve the performance. If there are extra RSUs, many locations are obtained for the same point due to the error of the positioning. Thus, the average must be taken by all the obtained locations to improve the accuracy.

- Each RSU knows its absolute location. Thus, the vehicles can determine its absolute location based on the estimated relative position and the absolute locations of the surrounding RSUs.

## 5 CONCLUSIONS AND FUTURE DIRECTIONS

In this paper, we have tackled the positioning problem for VANET applications. The proposed system is based on the use of the received signal strength (RSS) from several RSUs and the use of the DGPS.

For the use of the RSS, field experiments are conducted on realistic situations. The first one has a fatal error, which is the interference due to the use of the same wireless channel of all the WAPs. People who works on the same problem, uses this setup. Thus, their obtained results were not accurate. However, we have discovered the cause of this problem and how to solve it. This is manly by eliminating the effect of the channel interference and the near-far problems. In the second experiments, the results are very good compared to the first as the RSUs operated on different channels to reduce the interference. After analyzing the obtained results, we have proposed the use of the neural networks (NN) to learn the positioning problem from the obtained learning examples. The basic accuracy obtained is in the range of 60m for the first experiment (suffering from the interference). By using the data from the other experiments, the accuracy reached 8 meters (the maximum error). By using the curve fitting technique, and using only one RSU, we have done two experiments that show an obtained accuracy that can reach a maximum absolute error of only 2 meters. This can be obtained when the RSU is very far from the vehicle. The accuracy of determining the location can be less than 2m in the closed areas (tunnels, etc). We propose the use of DGPS in the areas where the GPS signal and DGPS correcting information can be obtained. In other areas, the proposed mechanism based on the RSS can be used.

In order to further improve the accuracy, we are currently exploring the use of nanotechnology device to determine the location precisely even in the areas where GPS signals cannot be received or very week.

## Authors

**Dr. Samir A. Elsagheer Mohamed** obtained a B.Sc. degree in Computer and Control Systems, from the Faculty of Engineering at the University of Assuit, Egypt, in May 1994. He worked as a teaching assistant in the Faculty of Engineering at Aswan from 1995 to 1997. He obtained his M.Sc. degree in Computer Science from the University of Rennes I, France, in 1998. He obtained his Ph.D. degree in Computer Engineering from the INRIA/IRISA, University of Rennes, France, in January 2003. Then, he worked as R&D Expert Engineer at the INRIA/IRISA until June 2006. After that he moved to the Faculty of Engineering at Aswan (Egypt) to work as assistant professor. Currently, he is with the College of Computer, Computer Engineering Department, at the Qassim University, Saudi Arabia. He is also the manager of the research unit in the College of Computer. In addition, he worked as a consultant for the IT center, Qassim University. His research interests are in vehicular ad hoc networks, ad hoc network security, audio and video quality assessment in computer networks; rate based control mechanisms; video codecs; sensor networks; traffic prediction; learning algorithms for neural networks; text classifications; cryptography; and road traffic safety.

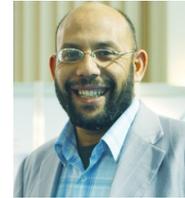

**Dr. Ashraf Nasr** received the Ph.D. and M.SC. degrees in optical communication engineering from Al-Azhar University, Cairo, Egypt, in 2004, and in electrical communication engineering from Menoufia University, Menouf, Egypt, in 1997, respectively. Presently, he is working in the College of Computer, Qassim University, Saudi Arabia. His current research interests are in nanotechnology applications in optical communications, optoelectronic devices, Ad-hoc carbon nanotube networks and solar cell. He has published more than 25 scientific papers in international journals.

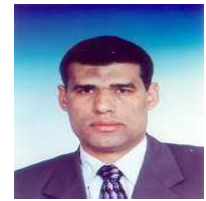

**Dr. Gufran Ahamd Ansari** received his Bachelor degree (B.Sc. Computer Science) from Shia P.G. College, Lucknow in 1997, Post graduate diploma from NIIT Lucknow, MCA from DR. B.R. Ambedkar University Agra in 2002 and Ph.D(Computer Science) from Babasaheb Bhimrao Ambedkar (A central) University, Lucknow, U.P., India in 2009. He is currently working as an Assistant Professor, Department of Computer Science, College of Computer , Qassim University, Kingdom of Saudi Arabia . Earlier he worked working as an Assiociate Professor Department of Computer Science & Engineering at MIT Meerut, lecturer at Azad Institute of Engineering & Technology (AIET) Lucknow, Lecturer, Senior Lecturer and Assistant Professor at Institute of Foreign Trade & Management (I.F.T.M), Moradabad U.P., India. He has more than 10 years of experience in teaching undergraduate as well as postgraduate students of Computer Science and Computer Applications. Currently he is actively engaged in the research work on domain based of Real-time system modeling through the Unified Modeling Language (UML). He has produced several outstanding publications in National & International Journal on various research problems related to the domain based UML modeling & Security, Testing and Designing.

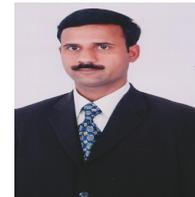